\documentclass[reprint,amsmath,amssymb,aps,hidelinks,pre]{revtex4-2}

\usepackage{graphicx}
\usepackage{float}
\usepackage{dcolumn}
\usepackage{bm}
\usepackage{hyperref}
\usepackage{orcidlink}
\hypersetup{colorlinks=true,linkcolor=blue,urlcolor=blue,citecolor=blue}
\usepackage{siunitx}

\usepackage{ulem}

\newcommand*{\figref}[2][]{%
  \hyperref[{fig:#2}]{%
    \ref*{fig:#2}%
    \ifx\\#1\\%
    \else
      #1%
    \fi
  }%
}

\begin{document}

\preprint{APS/123-QED}

\title{State-dependent complexity of the local field potential\\in the primary visual cortex}

\author{Rafael M. Jungmann\textsuperscript{1,\orcidlink{0000-0001-8952-8888}}  }
\author{Thaís Feliciano \textsuperscript{1,\orcidlink{0000-0002-3028-3094} } } 
\author{Leandro A. A. Aguiar \textsuperscript{1,2,3,\orcidlink{0000-0002-8195-4601}} }
\author{Carina Soares-Cunha \textsuperscript{2,3,\orcidlink{0000-0001-9470-644X}} }
\author{B\'arbara Coimbra \textsuperscript{2,3,\orcidlink{0000-0003-1737-2268}}  }
\author{Ana Jo\~ao Rodrigues \textsuperscript{2,3,\orcidlink{0000-0003-1968-7968}} }
\author{Mauro Copelli \textsuperscript{1,\orcidlink{0000-0001-7441-2858}} }
\author{Fernanda S. Matias \textsuperscript{5,\orcidlink{0000-0002-5629-6416}} }
\author{Nivaldo A. P. de Vasconcelos\textsuperscript{2,3,4,\orcidlink{0000-0002-0472-8205} } } \email{nivaldo.vasconcelos@ufpe.br}
\author{Pedro V. Carelli \textsuperscript{1,\orcidlink{0000-0002-5666-9606}}} \email{pedro.carelli@ufpe.br}

\affiliation{\textsuperscript{1}Departamento de Física, Universidade Federal de Pernambuco, 50670-901. Recife-PE, Brazil}
\affiliation{\textsuperscript{2}Life and Health Sciences Research Institute (ICVS), School of Medicine, University of Minho, Braga, 4710-057, Portugal}
\affiliation{\textsuperscript{3}ICVS/3B’s - PT Government Associate Laboratory, Braga/Guimarães, Portugal}
\affiliation{\textsuperscript{4}Departamento de Engenharia Biomédica, Universidade Federal de Pernambuco. 50670-901. Recife-PE, Brazil}
\affiliation{\textsuperscript{5}Instituto de Física, Universidade Federal de Alagoas. 57072-970. Maceió-AL, Brazil}

\date{\today}

\begin{abstract}
  
The local field potential (LFP) is as a measure of the combined activity of neurons within a region of brain tissue. While biophysical modeling schemes for LFP in cortical circuits are well established, there is a paramount lack of understanding regarding the LFP properties along the states assumed in cortical circuits over long periods. Here we use a symbolic information approach to determine the statistical complexity based on Jensen disequilibrium measure and Shannon entropy of LFP data recorded from the primary visual cortex (V1) of urethane-anesthetized rats and freely moving mice. Using these information quantifiers, we find consistent relations between LFP recordings and measures of cortical states at the neuronal level. More specifically, we show that LFP's statistical complexity is sensitive to cortical state (characterized by spiking variability), as well as to cortical layer. In addition, we apply these quantifiers to characterize behavioral states of freely moving mice, where we find indirect relations between such states and spiking variability.

\end{abstract}

\maketitle

\section{Introduction} \label{sec:Introduction}

Cortical neurons exhibit remarkable variability in their spiking patterns~\cite{Shadlen1998}. The technological advance of simultaneous recording techniques for large neuronal populations~\cite{Nicolelis2003,Csicsvari2003,Stevenson2011,Jun2017} has revolutionized our understanding of cortical dynamics, particularly the variability of spiking activity in several cortical areas, including the primary visual cortex. This spiking variability during spontaneous periods, exhibits similarities to the spiking activity triggered by stimuli in freely moving animals.~\cite{Tsodyks1999,Ringach2009}. Notably, the level of spiking variability in primary sensory cortices varies across different cortical states~\cite{Renart2010,Harris2011,McGinley2015,Scholvinck2015}, suggesting that it serves as a proxy for cortical state at both macro and microscopic scales. This variability appears to be a dynamic property, changing over time on a timescale of seconds. Therefore, investigating the state-dependence of cortical functions is fundamental for understanding the intricate mechanisms underlying cortical computations and their modulation by multiple behavioral factors.

The power spectrum of combined electrophysiological activity in the cortical extracellular medium typically exhibits a $1/f$ structure, characterized by elevated power in lower-frequency components~\cite{henrie2005,Buzsaki2012}. These components are primarily characterized by the concerted local activity of neural networks, which takes place within a restricted volume of tissue surrounding the electrode tip~\cite{Buzsaki2012}. Such time-varying signal has been termed local field potential (LFP). The use LFP to assess the overall neural function within a specific tissue area is gaining more and more attention~\cite{Mitzdorf1985,Buzsaki2012,Katzner2009}. LFP is mainly believed to reflect the synaptic currents of neuron populations in the vicinity of the recording site, being sensitive to the geometry and arrangement of the neurons~\cite{Herreras2016}. However, any transmembrane current in the brain tissue contributes to the LFP, giving rise to limitations in the interpretation of the signal's precise origin and spatial reach~\cite{Katzner2009,Linden2011,Kajikawa2011}. Despite this hardship, analyzing LFP signals is essential to understanding presynaptic activity and inferring properties of network dynamics. However, there is a lack of studies regarding the complexity of LFP signals along different cortical states, mainly in the sensory areas. 

To address this issue, we used simultaneous recordings of large neuronal populations in the primary visual cortex, in multiple experimental conditions: deep layers of urethane-anesthetized rats, and all layers of freely moving mice. Based on these data, segmented by level of variability of population spiking activity, we employ two quantifiers based on Information Theory: Shannon entropy, and MPR-statistical complexity, based on the disequilibrium between the actual time series and one with a uniform probability distribution function~\cite{Shannon1949,Lamberti2004,Rosso2007,Zunino2012,Xiong2020,Lopez1995}. We assign to each system under study a position in a two-dimensional space spanned by an entropy and a statistical complexity measure. These quantifiers are evaluated using the probability distribution function (PDF) obtained with the Bandt-Pompe symbolization methodology~\cite{Bandt2002}.

This approach was originally introduced to distinguish chaotic from stochastic systems~\cite{Rosso2007} in time series analysis. Recently, it has been applied to different brain signals:  to show that complexity is maximized close to criticality in cortical states using spiking data~\cite{Lotfi2021}, to distinguish cortical states using EEG data~\cite{Rosso2006}, MEG data~\cite{Mendoza2020}, as well as neuronal activity~\cite{Montani2015_2,Montani2014,DeLuise2021}. Furthermore, it has been applied to monkey LFP to estimate response-related differences between Go and No-Go trials~\cite{Lucas2021}, to estimate time differences during phase synchronization~\cite{Montani2015_1}, and to explore Hénon maps as a model for brain dynamics based on LFP data from subthalamic nucleus (STN) and medial frontal cortex (MFC) of human patients~\cite{Guisande2023}.

\section{Methods} \label{sec:Methods}

\subsection{Surgery and recordings}

The \textit{in vivo} experiments targeted recordings in albino urethane-anesthetized rats ($n = 4$, Wistar Han, male, 350-500 g, 3-6 months old, 1.44 g/kg urethane), from the primary visual cortex (V1, Bregma: AP = -7.2 mm,  ML = 3.5 mm). Such as is illustrated in Fig.~\figref[(a)]{tabela1}, the recordings has been done using a 64-channel silicon probe, with six shanks, 200 $\mu$m apart, inserted around the central coordinate, along the direction defined by the anterior-posterior axis~\cite{Paxinos2016}. The extra-cellular signal in each channel was sampled at 30 kHz (Intan RHD2164, 16 bits/sample), based on Open Ephys acquisition system~\cite{Siegle2017}. The \textit{in vivo} experimental procedures, encompassing animal housing, surgical interventions, and data recordings, strictly adhered to the FELASA guidelines~\cite{Nicklas2002} and were conducted in full compliance with European Regulations (European Union Directive 2010/63/EU). Both the animal facilities and the personnel responsible for conducting these experiments were officially certified by the Portuguese regulatory body, DGAV (Direcção-Geral de Alimentação e Veterinária). Furthermore, all research protocols underwent thorough scrutiny and received approval from the Ethics Committee of the Life and Health Sciences Research Institute (ICVS). More detailed information can be found in our recent studies~\cite{Vasconcelos2017,Fontenele2018}. Additionally, we employed non-albino urethane-anesthetized Long-Evans male rats ($n = 3$, 350--500 g, 3--6 months old, 1.44 g/kg urethane), under approval by the Federal University of Pernambuco (UFPE) Committee for Ethics in Animal Experimentation (23076.030111/2013-95, 12/2015, and 20/2020). Additionaly, we used recordings from 19 freely moving mice, implanted with 64-site linear silicon  probe  (H3, Cambridge NeuroTech). The freely moving mice database is public from Buzsáki's lab, as recently published~\cite{Senzai2019}. In summary, surgery and electrode implantation, extracellular electrophysiological recording, electrolytic lesions, histological processing, spike sorting, and detection of monosynaptic functional connectivities were performed . Electrophysiological data were acquired using an Intan RHD2000 system, digitized with a 20 kHz rate, and the wide-band signal was downsampled to 1.25 kHz for use as the LFP signal. 

\subsection{Data analysis}

The basic neuronal spiking data is modelled as a set of spike trains, where each spike train can be defined as:

\begin{equation}
    s_k(t) = \sum_{t_i \in T_k}{\delta(t-t_i)},
    \label{eq.spike_train}
\end{equation}
where $T_k$ is a list with the spike times of the $k$-th neuron within the neuronal population. Based on the Eq.~\ref{eq.spike_train}, the corresponding \textit{instantaneous firing rate} is defined as:

\begin{equation}
    r_k(t, \Delta t)=\frac{1}{\Delta t} \int_t^{t+\Delta t} s_k(\tau) d \tau,
    \label{eq.ifr}
\end{equation}
where $\Delta t$ defines the time resolution of this measure. Once one selects the time resolution, $\Delta t$ (also known as the bin size), the result from the Eq.~\ref{eq.ifr} is discrete and can represented by a corresponding time-series, $r_{k,i}$, with approximation based on a rectangular kernel ($\Delta t$-long). Based on the population instantaneous firing rate, on its discrete form, $r_{i}$, we calculated the corresponding \textit{coefficient of variation} as:

\begin{equation}
    CV_{i} = \frac{\sigma_{i}}{\mu_{i}},
    \label{eq.CV}
\end{equation}
where, numerator and denominator are standard-deviation and average values for the $i$-th $10$-s-long population instantaneous firing rate time-series, respectively. For each $10s$ period, the cortical state has been assessed by using the corresponding coefficient of variation of population instantaneous firing rate time-series, according to Eq.~\ref{eq.CV}.

In total, single-unit and multi-unit spiking activity encompassed 2535 neurons (833 from albino rats, 921 from non-albino rats and 781 from freely moving mice). The LFP recorded from all animals was tracked using a 500 Hz sample rate and a low-pass Butterworth filter, with a cutoff frequency of 200 Hz, was applied. In total, we used $6000s$-long data for each rat and $8000s$-long data for each mouse.

\subsection{Information quantifiers}
 
Let $X(t) \equiv \{ x_t;t=1,2,...,M \}$ be the time series representing a set of $M$ measurements of the observable $X$. We can associate to $X$, a probability distribution given by $P \equiv \{ p_j;j=1,2,...,N \}$ where $\sum_{j=1}^{N}p_j = 1$ and $N$ is the number of possible states of the system. given the PDF

After evaluating the PDF ($P \equiv \{ p_j;j=1,2,...,N \}$) associated to a time series $X(t)$ one can calculate different information quantifiers such as entropy and complexity. The Shannon's entropy associated to the distribution $P$ is defined by 

\begin{equation}
    S[P] = - \sum_{j=1}^{N} p_j \ln(p_j).
\end{equation}

This function is a information measure which is equal to zero when we can predict with certain the outcome of the observable. By contrast, the entropy is maximized when all states have equal probability (uniform distribution) $P_e = \{ p_j=1/N, \forall j = 1,2,...,N\}$. Therefore, the normalized Shannon's entropy is defined by $H[P] = S[P]/S[P_e]$ where $0 \leq H \leq 1$.

Here we consider the Martín-Platino-Rosso statistical complexity (MPR) \cite{Martin2006}, based on the notion of a statistical complexity proposed by López-Ruiz et al.~\cite{Lopez1995}. The main idea of this complexity measure is to differentiate systems on intermediate configurations between complete order ($H=0$) and disorder ($H=1$). These opposite extremes of perfect order and maximal randomness are too simple to describe and the complexity should be zero in both cases. 

Therefore, the complexity can be defined through the product
\begin{equation}
    C[P] = Q_j[P,P_e]\cdot H[P],
\end{equation}
where $Q_J[P,P_e]$ is a disequilibrium defined in terms of the Jensen-Shannon divergence as

\begin{equation}
    Q_j[P,P_e] = Q_0 J[P,P_e],
\end{equation}
where

\begin{equation}
    J[P,P_e] = S \left[ \frac{(P+P_e)}{2} \right] - \frac{S[P]}{2} - \frac{S[P_e]}{2}, 
\end{equation}
and $Q_0$ is a normalization constant ($0 \leq Q_J \leq 1$). $Q_0$ is equal to the inverse of the maximum value of $J[P,P_e]$, which is obtained when one of the $N$ states, say the state $m$, has $p_m = 1$ and the remaining states have $p_{j \neq m} = 0$. The Jensen-Shannon Divergence $J[P,P_e]$ is used to quantify the difference between the distribution associated with the system of interest and the uniform distribution. 

It has been demonstrated that, for a given value of normalized entropy $H$, the complexity $C$ can vary between a well-defined minimum $C_{min}$ and a maximum $C_{max}$ value which restricts the possible occupied region in the $C-H$ plane \cite{Martin2006}. 

\subsection{Bandt and Pompe symbolization technique}

Here, we use a symbolic representation of a time series introduced by Bandt and Pompe~\cite{Bandt2002} for evaluating the probability distribution function (PDF) associated to each time series $X(t)$ of interest.  This symbolization technique consist of extract the ordinal patterns of length $D$, by  indexing each time $t$ to the $D$-dimensional vector $\textbf{s}(t)=(x_{t},x_{t+1}, \cdots ,x_{t+D-1},x_{t+D})$.

The specific $j-$th ordinal pattern associated to $\textbf{s}(t)$ is the permutation $\pi_j=(r_{0},r_{1},...,r_{D-1})_j$ of $(0,1,...,D-1)$ which guarantees that $x_{t+r_{0}} \leqslant x_{t+r_{1}} \leqslant  \cdots  \leqslant x_{t+r_{(D-2)}} \leqslant x_{t+r_{(D-1)}}$. In other words, each permutation $\pi_j$ (with $j=1,2,...,D!)$ is one of our possible symbols and we have $D$! different symbols (ordinal patterns). To calculate the PDF we should identify and count the number of occurrences of each symbol $\pi_j$ of length $D$.

This procedure is essential to a phase-space reconstruction with embedding dimension (pattern length) $D$. For practical purposes, Bandt-Pompe \cite{Bandt2002} suggested to use $ 3\leqslant D \leqslant 7 $. Note that the probabilities to evaluate the PDF naturally arises from the time series after defining the symbols. This technique takes into account the temporal structure of the time series and yields information about the temporal correlation of the system.

To have an example, choosing $D=3$, all the 6 possible symbols associated with the permutations $\pi_j$ are: $\pi_1=(0,1,2)$, $\pi_2=(0,2,1)$, $\pi_3=(1,0,2)$, $\pi_4=(1,2,0)$, $\pi_5=(2,0,1)$, $\pi_6=(2,1,0)$. Considering an illustrative time series $X(t)=\{2,7,4,1,3,6,0,8,5\}$, the first vector is $\textbf{s}(t=1)=(2,7,4)$, corresponding to the permutation $\pi_2=(0,2,1)$; the second vector is $\textbf{s}(t=2)=(7,4,1)$, corresponding to to the permutation $\pi_6=(2,1,0)$; the third vector is $\textbf{s}(t=3)=(4,1,3)$, corresponding to the permutation $\pi_5=(2,0,1)$. Similarly, one can find the other 4 vectors $\textbf{s}(t)$ and its respective $\pi_j$ in order to build the PDF associated to $X(t)$. 

It is also possible to include a time delay $\tau$ to evaluate the PDF in different time scales. This means that we can skip every $\tau-1$ points of our time series $X(t)$ in order to find and count the symbols. In the above example we use $\tau=1$ and consider every point in $X(t)$. For $\tau=2$ we would skip every other point, in such a way that the first vector is $\textbf{s}(t=1)=(2,4,3)$, corresponding to the permutation $\pi_2=(0,2,1)$; the second vector is $\textbf{s}(t=2)=(7,1,6)$, corresponding to the permutation $\pi_5=(2,0,1)$; the third vector is $\textbf{s}(t=3)=(4,3,0)$, corresponding to the permutation $\pi_6=(2,1,0)$. For $\tau=3$ the first vector is $\textbf{s}(t=1)=(2,1,0)$, the second vector is $\textbf{s}(t=2)=(7,3,8)$; the third vector is $\textbf{s}(t=3)=(4,6,5)$. Therefore, to each time series $X(t)$ we can associate many PDFs, each one for a different value of the time delay $\tau$. 

\begin{figure*}[!htb]
    \centering
    \includegraphics[width=2.0\columnwidth]{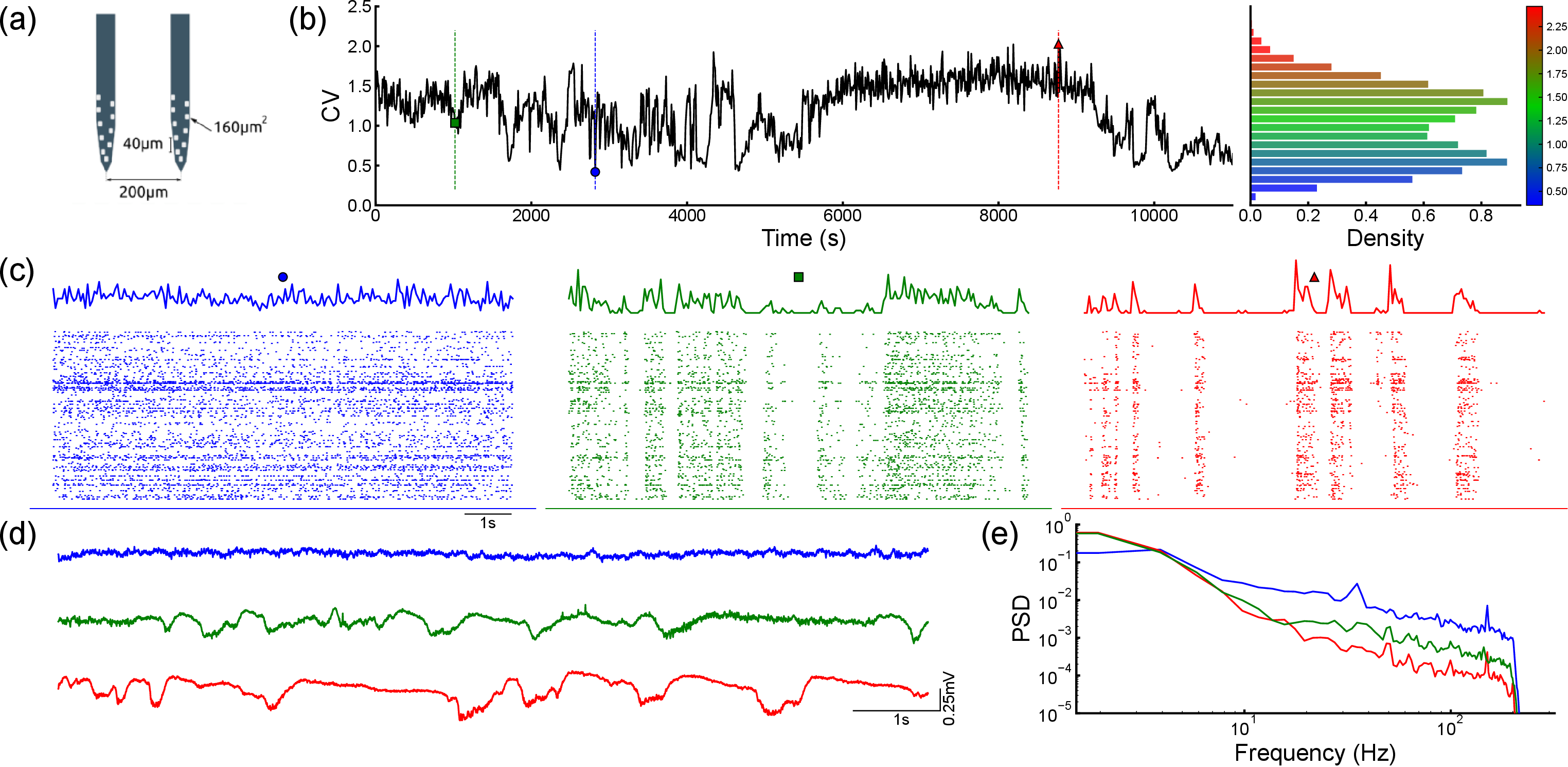}
    \caption{Cortical dynamics in different levels of spiking variability. (a) Schematics of silicon probe (6 shanks) inserted in the rat's primary visual cortex (V1). (b) Coefficient of variation (CV) of spiking activity calculated for $10$-s-long non-overlapping windows, followed by the CV histogram of a single animal; Symbols (blue circle, green square, and red triangle) indicate the level of spiking variability of three representative examples: low (CV$=0.42$, desynchronized activity), intermediate (CV$=1.04$), and high (CV$=2.03$, highly synchronous activity). (c) Spiking activity across the three levels of CV depicted in (b). (Top) Firing rate calculated using $50$ ms bin. (Bottom) Raster plots of single- and multi-units activity (SUA$=90$ and MUA$=145$). (d) $10$-s-long LFP data across the three levels of CV. LFP is tracked using $500$ Hz sample rate and a low-pass Butterworth filter is applied with cutoff frequency of $f_{cutoff}=200$ Hz. (e) Power spectrum density (PSD) of data shown in (d).}
    \label{fig:tabela1}
\end{figure*}

\section{Results} \label{sec:Results}
\subsection{Local neuronal populations in deep layers of V1} \label{sec:Anesthetized rats}

\begin{figure*}
    \centering
    \includegraphics[width=1.75\columnwidth]{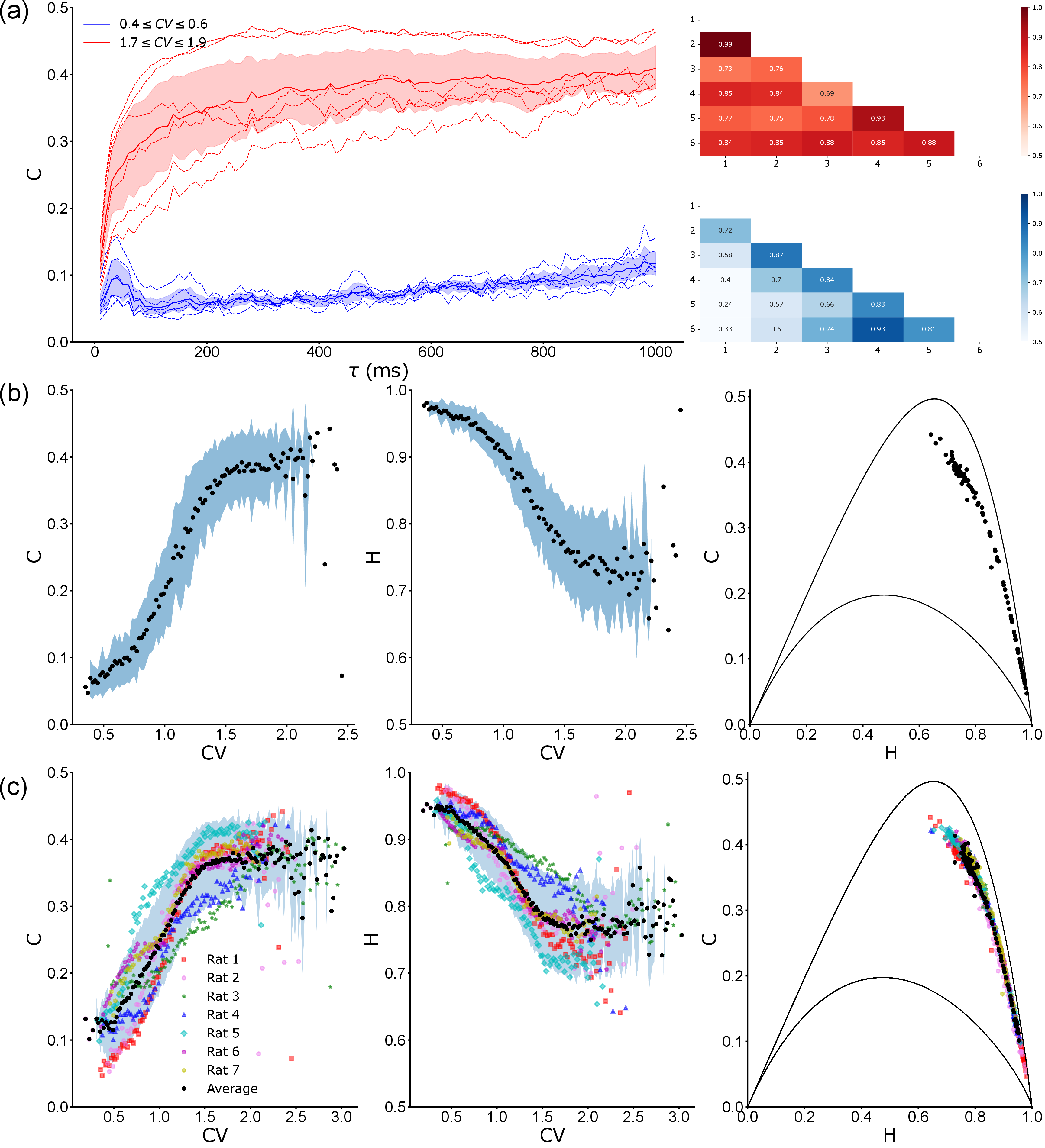}
    \caption{Complexity and entropy quantifiers across different levels of spiking variability. (a) (left) Statistical complexity of $10$-s-long windows of LFP data versus the time delay $\tau$ calculated for each shank (dashed lines) for low (blue) and high (red) CV. Average (lines) and standard deviation (shading) is performed over all shanks. (right) Complexity correlation matrix of all $6$ shanks for low and high CV. (b) Statistical complexity (left) and Shannon entropy (middle) of LFP versus CV of the spiking activity for $\tau = 200$ ms. Group average (black circles) and standard deviation (blue shading) is taken over all shanks using bins of $0.02$ for the CV axis. Complexity and entropy plane (right) is represented using group average. Black lines represent the theoretical maximum $C_{max}$ and minimum $C_{min}$ complexity values in the $C-H$ plane for $D=6$. (c) Group data for corresponding plot in (b), from primary visual cortex of 7 animals, in 42 local neuronal populations.}
    \label{fig:tabela2}
\end{figure*}

Deep layers of pyramidal neurons in rodents' primary visual cortex have dendrites reaching all six layers, making them important integrators in the cortical column~\cite{Shai2015}.  We already investigated its complexity based on its spiking activity~\cite{Lotfi2021}. However, an analysis of the statistical complexity of local field potential in deep layers of V1 is still lacking, especially when observed along multiple cortical states. Figure~\figref[(a)]{tabela1} displays an illustration of details for a pair of shanks of the silicon probe (Buzsaki64sp, Neuronexus) used to record six local neuronal populations, \SI{200}{\micro\metre} apart.

Here we associate different cortical states with different levels of summed spiking variability in large neuronal populations~\cite{Renart2010,Harris2011,Vasconcelos2017,Fontenele2018,Lotfi2020,Lotfi2021}. 
Figure~\figref[(b)]{tabela1} displays the time-series of the coefficient of variation (CV, Eq. \ref{eq.CV}), calculated over non-overlapping $10$-s-long windows, illustrating the summed spiking variability within neuronal populations in V1 of urethane-anesthetized rats. In addition, its right panel displays the corresponding CV histogram. Fig.~\figref[(c)]{tabela1} illustrates the detailed spiking activity across the same neuronal population in its different levels of summed spiking variability, according to the corresponding colored markers in (b): low (blue), intermediate (green), high (red). The top panels in 
Fig.~\figref[(c)]{tabela1} show the
firing rate calculated in a time window of 50 ms. The bottom panels show the raster plots of 90 single-units and 145 multi-units activity in the vertical axis. Each dot represents a spike.
Fig.~\figref[(d)]{tabela1} shows samples of $10$-s-long local field potentials along the different levels of summed spiking variability shown in (b) (using the same color code), whereas the Fig.~\figref[(e)]{tabela1} shows the corresponding power spectrum density (PSD) of LFP signals. At low CV, the spiking activity is desynchronized and the LFP power spectrum is relatively flat. At intermediate CV, the spiking activity becomes more synchronized. At high CV, the spiking activity is highly synchronous and the LFP power spectrum is dominated by low frequencies.

Recently, we used the statistical complexity and Shannon entropy analysis to investigate the signaling \textit{output} of deep layers in the primary visual cortex of urethane-anesthetized rats~\cite{Lotfi2020}: its spiking data. A natural step further in the study of different cortical states is to investigate the statistical complexity and Shannon entropy of LFP across multiple cortical states. Figure~\figref[(a)]{tabela2} shows the statistical complexity of $10$-s-long windows of LFP data versus the time delay $\tau$ calculated for each shank (dashed lines) for multiple levels of variability in population spiking activity (low CV in blue and high CV in red). Average (lines) and standard deviation (shading) is performed based on all shanks. The complexity correlation matrix of all 6 shanks for low and high CV, which measures the correlation between the time series of the statistical complexity calculated in different shanks of the same experimental setup, is also shown. Figure~\figref[(b)]{tabela2} shows the statistical complexity (left) and Shannon entropy (middle) of LFP versus CV of the spiking activity for  $\tau = 200$ ms. The average group data (black circles) and standard deviation (blue shading) are taken over all shanks of a single animal using bins of $0.02$ for the CV axis. Complexity and entropy plane (right) are represented using group average. Black lines represent the theoretical maximum $C_{max}$ and minimum $C_{min}$ complexity values in the $C-H$ plane for $D=6$. The results show that the statistical complexity of the LFP increases with increasing CV, regardless of the time delay $\tau$. This suggests that the LFP becomes more regular, less noisy (as expected for more synchronized regimes) with increasing spiking variability. The complexity correlation matrix also shows that the LFP signals across different shanks become more correlated with increasing spiking variability. The Shannon entropy of the LFP decreases with increasing CV, although the effect is less pronounced than for the statistical complexity. This suggests that the LFP becomes more predictable with increasing spiking variability.

The diverse group data based on all urethane-anesthetized rats, male/female, albino/non-albino (please refer to methods for more details) is shown in Fig.~\figref[(c)]{tabela2}. Group average (black circles) and standard deviation (blue shading) are taken over all 42 shanks of all 7 rats using bins of $0.02$ for the CV axis. The average of each animal is also presented (colored markers). Qualitatively, results are the same for each rat: increasing relation between the statistical complexity of the LFP and CV (left) and inverse relation between the Shannon entropy of the LFP and CV (middle). The $C-H$ plane (right) suggests the maximum values achieved of complexity are associated with the minimum values achieved of entropy which both occur for higher spiking variability (CV $\ge 1.5$). Overall, the results suggest that the statistical complexity and Shannon entropy of the LFP are both sensitive to changes in spiking variability. This suggests that these measures could be used to assess changes in the dynamics of cortical circuits.

\subsection{Statistical complexity in layers} \label{sec:Layers}

\begin{figure*}
    \centering
    \includegraphics[width=1.75\columnwidth]{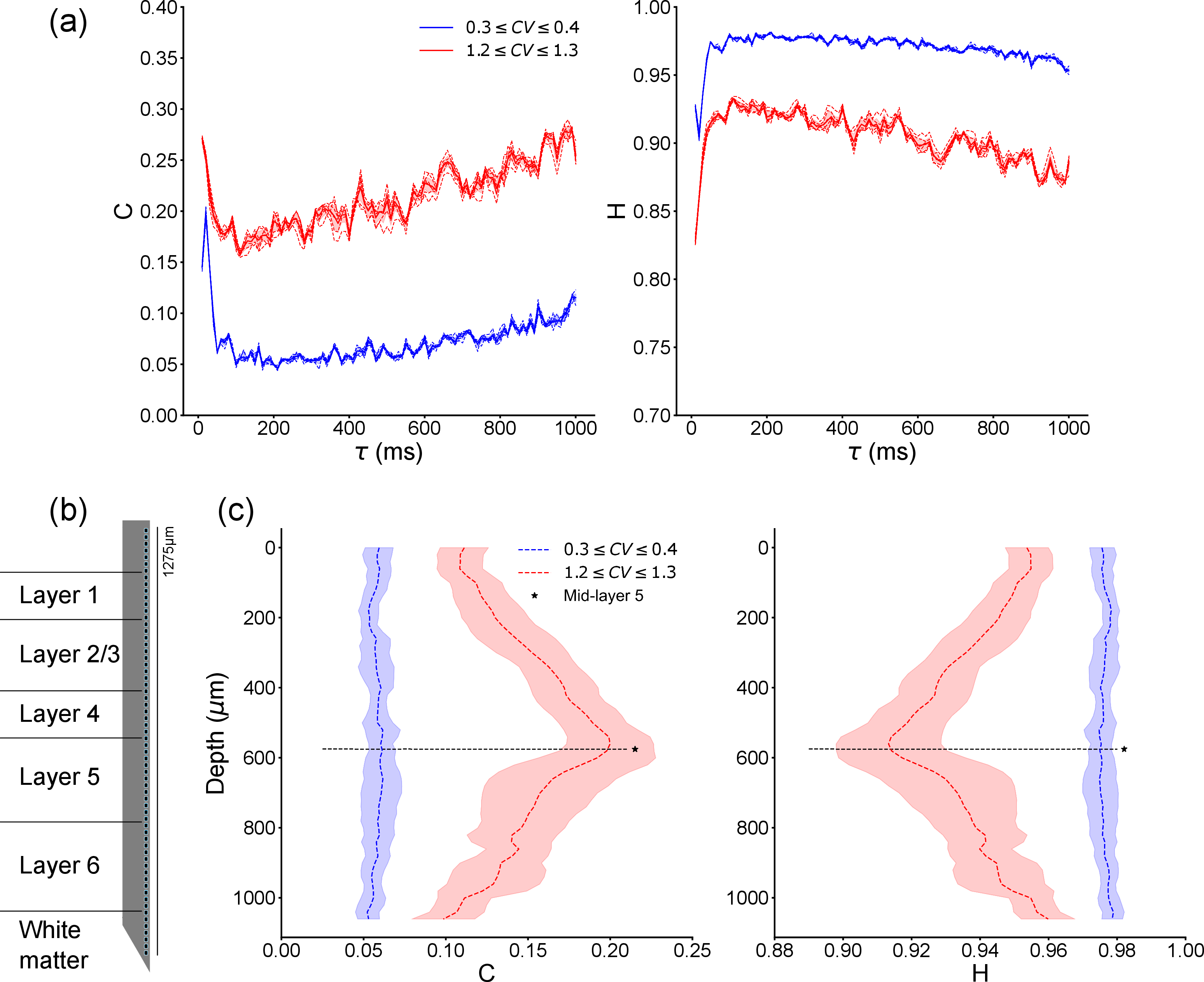}
    \caption{Depth profile of complexity and entropy quantifiers across different levels of spiking variability. (a) Statistical complexity and entropy of $10$-s-long windows of LFP data versus the time delay $\tau$ calculated for different channels (dashed lines) for low (blue) and medium (red) CV. Average (lines) and standard deviation (shading) is performed over close by ($\approx 100$ $\mu$m) channels in the mouse's V1. (b) Vertical silicon probe illustration. It contains $64$ channels with $20$ $\mu$m gaps between each site. (c) Complexity and entropy of LFP data versus normalized depth (layer 1 is marked as zero for each mouse) for low (blue) and medium (red) CV for $\tau = 200$ ms. Quantifier's average (colored dashed lines) and standard deviation (shading) is calculated using all 7 mice. Mid-layer 5 (black star) is represented as an average over each mouse.}
    \label{fig:tabela3}
\end{figure*}

Previous studies have shown that the laminar structure of spontaneous and sensory-evoked population activity in mammals primary sensory cortex are related to the local information processing and the flow of information through cortical circuits~\cite{Sakata2009,Senzai2019,Liu2022}.
Therefore, we also calculated the statistical quantifiers along the laminar axis (dorsal-ventral axis) for freely moving mice.

 Figure~\ref{fig:tabela3} illustrates the depth profile of complexity and entropy quantifiers of the presynaptic activity along the cortical states in the primary visual cortex of freely moving mice. As in the previous subsection, the cortical states were characterized by the level of spiking variability on the local summed population activity using the CV. In Fig.~\figref[(a)]{tabela3}), we examined the statistical complexity and entropy of $10$-second-long windows of local field potential data as a function of the time delay $\tau$. We calculated these measures across different channels, denoted by dashed lines, for two distinct conditions: low spiking variability, represented by the blue lines, and intermediate spiking variability, depicted by the red lines.
 
 To ensure statistical robustness, we average the results (solid lines) and provide the standard deviation of the mean (shaded areas) over nearby channels (approximately 100 µm) within V1. Figure~\figref[(b)]{tabela3} provides a visual representation of the vertical silicon probe used to record the neuronal data from freely moving mice~\cite{Senzai2019}. The adoption of such a probe provided us with LFP data at varying depths in the mouse's V1, and we analyzed the corresponding complexity and entropy measures along these depths. In Fig.~\figref[(c)]{tabela3}, we focus on the relationship between the complexity and entropy of LFP data and their normalized depth. We adopted $D=6$, and $\tau$ = \SI{200}{\milli\second} and again considered low (blue) and medium (red) variability conditions. The dashed lines represent the average quantifier values, while the shaded regions indicate the standard deviation of the mean. Importantly, our analysis encompasses data from seven different mice, allowing us to derive general trends. We particularly highlight the significance of mid-layer 5 (denoted by the black star), which is presented as an average across all mice, shedding light on the neural dynamics at this specific depth. 

Our findings indicate that, for larger time delays ($\tau > 100$~ms), there is a clear distinction in the statistical complexity and entropy of freely moving mice LFP under multiple levels of spiking variability, as in the case of anesthetized rats. In addition, a maximization of the statistical complexity is found in mid-layer 5 as we increase CV. The entropy appears to be the minimum for this depth when the CV is increased.  

\subsection{Statistical complexity across behavioral states}

\begin{figure}
    \centering
    \includegraphics[width=1.0\columnwidth]{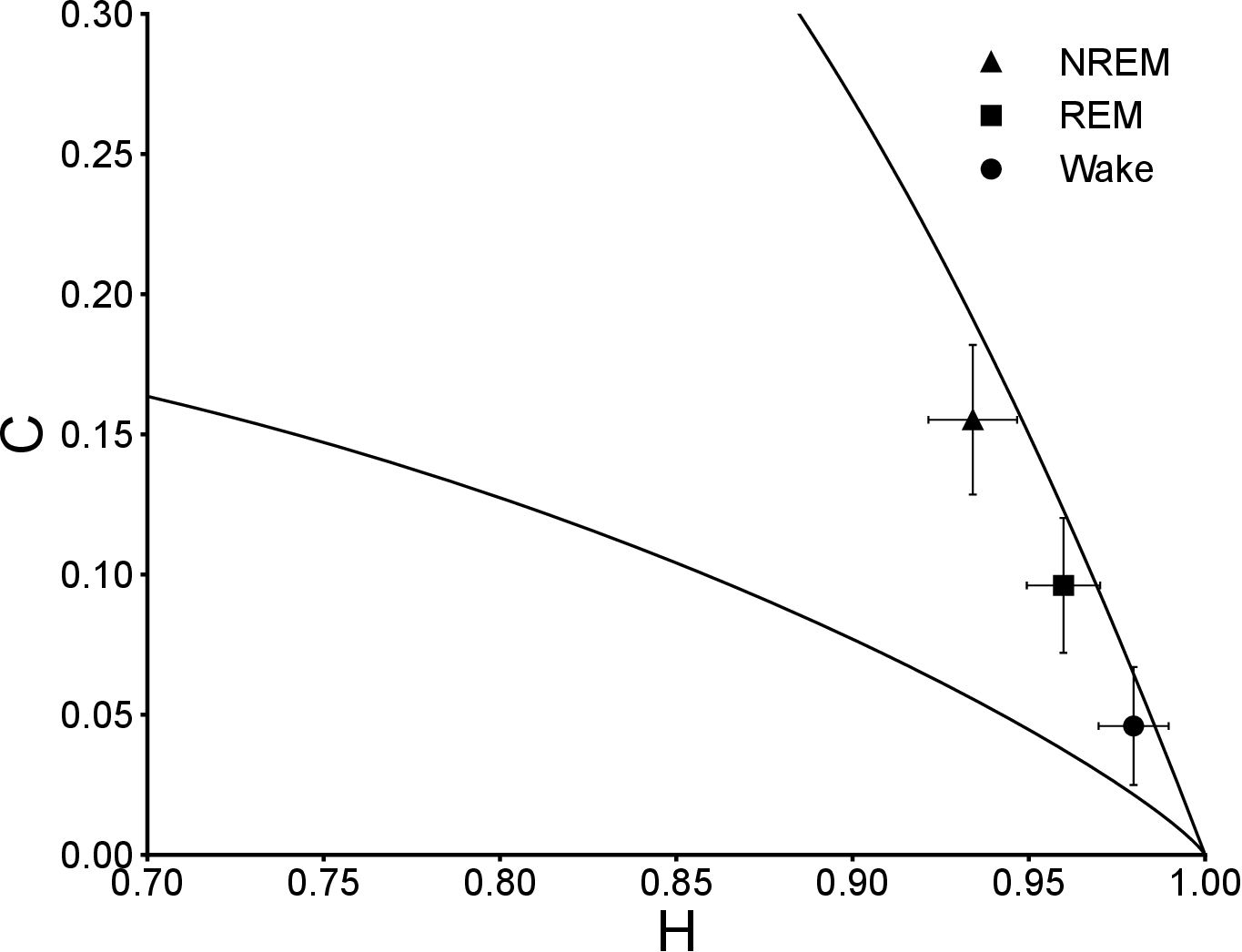}
    \caption{Entropy-complexity plane $C-H$ of V1's layer 5 LFP data along different behavioral states: NREM (triangle), REM (square) and awake (circle), where the marker indicates the corresponding mean value for each behavioral group in 10-s-long episodes of NREM, REM, and awake states for all 7 mice; where $D=6$ and $\tau = 200$ ms; both complexity and entropy values were significantly different among behavioral groups (p$\ll$ 0.01, Mann-Whitney test).}
    \label{fig:tabela4}
\end{figure}

Sensory responses can be influenced at both neural and behavioral levels~\cite{McGinley2015}. For example, in recent decades researchers delved into the intricacies of the wake-sleep cycle, a fundamental phenomenon exhibited by all higher vertebrates. They aimed to unravel the underlying neuronal mechanisms governing this cycle, employing intracranial local field potentials recorded from various brain regions, including the cortex, hippocampus, striatum, and thalamus~\cite{Gervasoni2004,Ribeiro2010}. Therefore, we also estimate the information quantifiers using behavioral states of the freely moving mice as parameters for data segmentation. Figure~\ref{fig:tabela4} shows the statistical complexity and entropy plane $C-H$ of V1's layer 5 LFP data along different behavioral states: NREM (non-rapid eye movement), REM (rapid eye movement), and awake which are represented by triangle, square, and circle markers, respectively. We have examined over 1900 NREM, 220 REM, and 1300 awake $10$-s-long episodes using $D=6$ and $\tau = 200$ ms, presenting the results as averages over all 7 mice. Figure~\ref{fig:tabela4} suggests that the statistical complexity of the LFP achieves higher values in NREM states and, in contrast, lower values when the mouse is awake. REM states, on the other hand, appear to be associated with intermediary values of complexity.

\section{Conclusions} \label{sec:Discussion}

 The current study fills a gap between LFP's quantifiers (complexity and entropy) versus a proxy of cortical state (coefficient of variation, CV, of the spiking activity).
 We have shown that LFP data can be characterized by information-theory quantifiers: Shannon entropy~\cite{Shannon1949} and Martín-Platino-Rosso statistical complexity~\cite{Martin2006}. We have employed a symbolic representation, based on Bandt-Pompe technique~\cite{Bandt2002}, to assign a probability distribution function to the LFP signals generated by urethane-anesthetized rats and freely moving mice.
 
The results of this study show that the statistical complexity and Shannon entropy of LFPs in deep layers of the primary visual cortex (V1) vary with the level of summed spiking variability. These results extend previous findings that have shown that the statistical complexity is a sensitive measure of cortical dynamics at the spiking level~\cite{Lotfi2021}. At low CV, the LFP is desynchronized, more noisy, and the power spectrum is relatively flat. As CV increases, the LFP becomes more synchronized and the power spectrum shows peaks at specific frequencies. At high CV, the LFP is highly synchronous and the power spectrum is dominated by a single low-frequency peak. We have shown that as CV increases the Shannon entropy of the LFP decreases and the complexity increases.
Therefore, these quantifiers could be potentially used as markers of cortical state. 

Recording LFP by inserting silicon probes in the brain allows us to not only explore electrical events at deeper layers, but along all layers. Therefore, the current study took advantage from the recent development regarding recordings from large neuronal population to quantify statistical complexity and entropy along the detailed laminar structure in the primary visual cortex~\cite{Sakata2009,Senzai2019}. We have shown that statistical complexity is sensitive to V1's depth for increasing CV, being maximized in mid-layer 5. This is particularly interesting, as it highlights the significance of such layer, which contains more branched neurons~\cite{Senzai2019}. 

Lastly, we have also shown that statistical complexity of mice LFP is significantly different across different behavioral states: NREM, REM and awake. The complexity is smaller for awake state, intermediate for REM sleep and larger for NREM sleep. Our statistical complexity results of mice LFP are similar to the ones reported in the analysis of sleep stages of EEG data recorded from humans~\cite{Mateos2021}. This findings opens new perspectives for using information theory quantifiers applied to LFP data to study cortical and behavioral states.

\section*{Acknowledgments}

We thank G. Buzs\'aki, Y. Senzai and P. Petersen for prompt support with the freely moving mice data.

R.M.J., T.F., L.A.A.A., M.C., N.A.P.V., and P.V.C. acknowledge support from CAPES (Grants No. 88887.131435/2016-00 and PROEX 534/2018, No. 23038.003069/2022-87), FACEPE (grant APQ-1187-1.05/22), and CNPq (Grants 142506/2018-4, 310712/2014-9, 308775/2015-5, 301744/2018-1, 425329/2018-6, 249991/2013-6, 408145/2016-1, 425037/2021-5, 311418/2020-1, and 308703/2022-7). F.S.M. acknowledge support from FAPEAL (grant APQ2022021000015) and CNPq (grant 314092/2021-8). C.S.-C. and B.C. acknowledge support from FCT (Grants No. SFRH/BD/51992/2012 and No. SFRH/BD/98675/2013) and PAC, MEDPERSYST Project POCI-01-0145-FEDER-016428 (Portugal 2020). A.J.R. received support from an FCT Investigator Fellow (IF/00883/2013) and acknowledges the Janssen Neuroscience Prize (first edition) and the BIAL Grant No. 30/2016.

\nocite{*}

\bibliographystyle{apsrev4-2}

\end{document}